\documentclass[runningheads]{llncs}

\usepackage{amsfonts}
\usepackage{amsmath}
\usepackage{amssymb}

\usepackage{multirow}
\usepackage{graphicx}
\usepackage{cite} 
\usepackage{color}
\usepackage{array}
\usepackage{tabularx}
\usepackage{longtable}
\usepackage{booktabs}
\usepackage[misc]{ifsym}

\begin{document}
\title{PARCEL: Physics-based Unsupervised Contrastive Representation Learning for Multi-coil MR Imaging} % Replace with your title
\titlerunning{Physics-based Unsupervised Contrastive Representation Learning}

\author{Shanshan Wang\inst{1,2,3,4,5}$^{(\textrm{\Letter})}$\and Ruoyou Wu \inst{1,2,3,4}\and Cheng Li\inst{1}\and Juan Zou\inst{1,6} \and Ziyao Zhang\inst{1,2,3}\and Qiegen Liu\inst{7}\and Yan Yi\inst{1} \and Hairong Zheng\inst{1}$^{(\textrm{\Letter})}$}
\authorrunning{S. Wang et al.}

\institute{Paul C. Lauterbur Research Center for Biomedical Imaging, Shenzhen Institutes of Advanced Technology, Chinese Academy of Sciences, Shenzhen 518055, China \and University of Chinese Academy of Sciences, Beijing 100049, China \and Peng Cheng Laboratory, Shenzhen 518055, China \and Guangdong Provincial Key Laboratory of Artificial Intelligence in Medical Image Analysis and Application, China \and National Center for Applied Mathematics Shenzhen (NCAMS), Shenzhen 518055, China \and School of Physics and Optoelectronics, Xiangtan University, Xiangtan 411105, China \and Department of Electronic Information Engineering, Nanchang University, Nanchang 330031, China
\\
\email{sophiasswang@hotmail.com, hr.zheng@siat.ac.cn}}

%******************
\maketitle

\begin{abstract}
With the successful application of deep learning to magnetic resonance (MR) imaging, parallel imaging techniques based on neural networks have attracted wide attention. However, in the absence of high-quality, fully sampled datasets for training, the performance of these methods is limited. And the interpretability of models is not strong enough. To tackle this issue, this paper proposes a Physics-bAsed unsupeRvised Contrastive rEpresentation Learning (PARCEL) method to speed up parallel MR imaging. Specifically, PARCEL has a parallel framework to contrastively learn two branches of model-based unrolling networks from augmented undersampled multi-coil k-space data. A sophisticated co-training loss with three essential components has been designed to guide the two networks in capturing the inherent features and representations for MR images. And the final MR image is reconstructed with the trained contrastive networks. PARCEL was evaluated on two vivo datasets and compared to five state-of-the-art methods. The results show that PARCEL is able to learn essential representations for accurate MR reconstruction without relying on fully sampled datasets. The code will be made available at https://github.com/ternencewu123/PARCEL. 
\end{abstract}

\section{Introduction}

PARALLEL imaging is an essential technique for accelerating MR imaging \cite{griswold2002generalized},\cite{pruessmann1999sense},\cite{sodickson1997simultaneous}. It utilizes magnetic resonance physics and the sensitivity of multiple coils to reconstruct MR images from multi-coil measurements either directly in the k-space domain \cite{griswold2002generalized} or in the spatial domain \cite{pruessmann1999sense}. For instance, Griswold et al. \cite{griswold2002generalized} proposed a partially parallel acquisition method, GRAPPA, to accelerate image acquisition. Pruessmann et al. \cite{pruessmann1999sense} proposed a sensitivity encoding (SENSE) method, and Sodickson et al. \cite{sodickson1997simultaneous} proposed a new fast-imaging technique (SMASH) to increase MR image acquisition speed. These methods have achieved great successes. However, the acceleration factor in classical parallel imaging is limited and its performance suffers from noise amplification effect \cite{robson2008comprehensive}. To address this issue, compressed sensing along with different sparse prior knowledges have been introduced into parallel MR imaging, which can better remove aliasing artifacts and suppress noise \cite{wang2017learning}, \cite{lustig2010spirit}, \cite{jin2016general}. However, it is difficult to determine the weight of these regularization terms in compressed sensing methods, and its inherent iterative reconstruction process is time-consuming \cite{hammernik2018learning}, \cite{yang2018admm}, \cite{wang2021deep}.

To further promote MR imaging speed and automate weight parameter settings, deep learning has been introduced to parallel MR reconstruction from undersampled data. These methods can be roughly divided into data-driven and model-based methods \cite{wang2021deep}. Data-driven methods require a neural network model to learn the mapping between the artifact images and high-quality images, or undersampled k-space data and fully sampled k-space data \cite{kwon2017parallel}, \cite{jun2019parallel}, \cite{akccakaya2019scan}, \cite{wang2016accelerating}, \cite{lee2018deep}, \cite{mardani2018deep}, \cite{sriram2020grappanet}, \cite{pawar2021domain}, \cite{feng2021dual}, \cite{feng2021donet}. For instance, Sriram et al. \cite{sriram2020grappanet} proposed GrappaNet, which integrate traditional parallel imaging methods into deep neural networks to generate high quality reconstructions. Pawar et al. \cite{pawar2021domain} incorporates domain knowledge of parallel MR imaging to augment the DL networks to achieve accurate and stable image reconstruction. Feng et al. \cite{feng2021donet} proposed a dual-octave (DONet) that can learn multi-scale spatial frequency features of MRI data, further accelerating parallel MRI reconstruction. Model-based methods based on the compressed sensing reconstruction algorithm unroll the iterative optimization process into a deep network, and use data training to optimize the parameters \cite{chlemper2017deep}, \cite{wang2020deepcomplexmri}, \cite{zhang2018ista}, \cite{aggarwal2018modl}, \cite{knoll2020deep}, \cite{chun2020momentum}, \cite{aggarwal2020j}. Generally, the model-based methods have better physical interpretability and is more robust compared to data-driven methods \cite{wang2021deep}. For example, Hammernik et al. \cite{hammernik2018learning} combined the variational model with deep learning and embedded compressed sensing reconstruction into the gradient descent method to achieve the rapid reconstruction of MR images. In addition, Aggarwal et al. \cite{aggarwal2018modl} proposed MoDL, which uses a deep neural network as a regularization term and the conjugate gradient algorithm to solve the inverse problem in combination with data consistency. Aggarwal et al. \cite{aggarwal2020j} proposed J-MoDL, which introduce a continuous strategy to optimize the sampling pattern and network parameters jointly. Multi-modal imaging is also an important branch of MR reconstruction \cite{feng2021multi}, \cite{feng2022multi}. Feng et al. \cite{feng2021multi} proposed a multi-stage integration network for super-resolution of multi-modal MR images.

The above methods have enabled great progress in accelerating MR imaging. However, they rely heavily on high quality, fully sampled MR images\cite{knoll2020deep}. To decrease the dependence on full reference data, self-supervised learning has been introduced for MRI \cite{yaman2020self}, \cite{liu2021magnetic}, \cite{yaman2021ground}, \cite{gan2021ss}, \cite{hu2021self}. Especially, Yaman et al. \cite{yaman2020self} proposed a self-supervised learning method (SSDU) for physics-guided deep learning reconstruction that divides measured data into two disjoint subsets, one used for training and the other used for loss function. This inspired method introduces the training of neural networks without fully sampled reference data. However, the performance of the method has room for improvements due to the under-utilization of the measurement data. They therefore further split the scanned undersampled measurements into multiple groups with each group consisting two sets of disjoint k-space data, for more effective deep learning MR reconstruction \cite{yaman2021ground}. In addition, Gan et al. \cite{gan2021ss} proposed a method named SS-JIRCS, which is a self-supervised model-based deep learning method for image reconstruction that is equipped with automated coil sensitivity map (CSM) calibrations. These methods have made encouraging contributions. Nevertheless, it is still worth investigation of unsupervised deep learning for parallel MR imaging. Contrastive representation learning \cite{chen2020simple} is a widely known unsupervised co-training framework that is very effective for obtaining essential and accurate representations for target samples. 

To induce higher reconstruction accuracy, we propose a physics-based unsupervised contrastive representation learning (PARCEL) method, which investigates and integrates the strengths of contrastive representation learning \cite{chen2020simple}, \cite{grill2020bootstrap}, \cite{bardes2021vicreg} and model-based deep learning MR reconstruction models \cite{aggarwal2018modl}. Specifically, we make the following key contributions in this study:

1) A PARCEL imaging framework is proposed which introduces unsupervised contrastive representation learning into parallel MR imaging. It simultaneously learns two model-based unfolding networks from augmented multi-coil k-space data and then uses them for the final accurate MRI reconstruction.

2) A sophisticated co-training loss with three essential components has been designed to guide the two networks in capturing the similarity of inherent features and representations and eliminating the information by which the two representations differ. Specifically, it has the undersampled calibration loss, reconstructed calibration loss and contrastive representation loss.

3) We compare PARCEL to five state-of-the-art methods with different sampling masks. The results show that PARCEL achieves good results in both qualitative and quantitative evaluations, which closely approaches the results achieved by supervised learning methods. In addition, PARCEL achieves better reconstructions than existing self-supervised methods.

The remainder of this paper is organized as follows. Section 2 introduces parallel magnetic resonance imaging and a brief recap of MoDL. Section 3 introduces the proposed method PARCEL. Section 4 summarizes the experimental details and results. Section 5 is the discussion part and Section 6 concludes the paper.

% 2 PRELIMINARY
% 2.1 Compressed Sensing based Parallel MR Imaging 
\section{Preliminary}
\subsection{Compressed Sensing based Parallel MR Imaging}

In parallel MR imaging, multiple receiver coils are used to accelerate MR imaging. Specifically, for the $i$-th coil measurement, we have:
% equa 1
\begin{equation}
    y_{i} = \mathbf{\Omega} FS_{i}x+\varepsilon_{i},\quad i = 1,2,...,C
\end{equation}
where $y_{i}\in \mathbb{C}^M$ represents the k-space measurement corresponding to the $i$-th coil, $\mathbf{\Omega} $ indicates the sampling mask, $\varepsilon_{i}\in \mathbb{C}^M$ represents measurement noise, $C$ represents the number of coils to be measured, $F$ is the normalized Fourier transform, $x\in \mathbb{R}^N$ is the to be reconstructed image, and $S_{i}$ represents the sensitivity map of the $i$-th coil. $S_{i}$ can be estimated using the k-space region corresponding to low frequencies (also known as the autocalibration signal, or ACS), which is fully sampled. In this experiment, we use the ESPIRiT \cite{uecker2014espirit} algorithm to obtain the sensitivity maps. When appling compressed sensing to parallel MR imaging, the minimization formula can be described as follows:
% equa 2
\begin{equation}
    x^{\ast}=\mathrm{arg}\min_{x}\frac{1}{2}\left \| Ax-y \right \| _{2}^{2}+\lambda \mathcal{R}(x)  
\end{equation}
where $A=\mathbf{\Omega} FS$ represents the encoding matrix with the diagonal matrix $S$ denoting the stack of all the coil sensitivities $S=diag\left \{S_{i}  \right \} $. The first term represents the data consistency term; $\mathcal{R}(x)$ represents the regularization term and $\lambda$ represents the regularization coefficient.

% 2.2 A Brief Recap of MoDL 
\subsection{A Brief Recap of MoDL}
In order to solve (2), the regularization term can be a denoising regularization. A typical example is MoDL proposed in \cite{aggarwal2018modl}, which attempts to solve the following optimization problem:
% equa 3
\begin{equation}
    x^{\ast}=\mathrm{arg}\min_{x}\frac{1}{2}\left \| Ax-y \right \| _{2}^{2}+\lambda \left \| N_{w}\left ( x \right )  \right \|^{2}
\end{equation}
where $\lambda$ represents a trainable regularization parameter and $N_{w}\left (x \right ) = x-D_{w}\left ( x \right ) $ denotes a learned convolutional neural network (CNN) of noise and alias corresponding to the "denoised" version $D_{w}\left ( x \right ) $ of $x$. With the alternating minimization algorithm, the model was solved as the following iteration process:
% equa 4
\begin{equation}
\left\{
    \begin{array}{lr}
    z^{k}=D_{w}\left (x^{k}  \right ) &  \\
    x^{k+1}=\left (A^{H}A+\lambda I  \right ) ^{-1}\left (A^{H}y + \lambda z^{k}\right )
    \end{array}
\right.
\end{equation}
where $k$ is the iteration number and $z^{k}$ is the intermediate denoised version of the image $x^{k}$. This iteration process is unrolled into a fully supervised learning process, which has two main modules, namely the residual learning based denoiser module $z=D_{w}(x)$ and the data consistency module for updating the image $x^{k}$. Here, the data consistency constraint is solved by conjugate gradient method. Fig. 1 shows the specific MoDL architecture. The CNN based denoiser block $D_{w}(x)$ is shown in Fig. 1(a) and the unrolled neural network architecture is shown in Fig. 1(b), whose weights are shared at different iterations.

% MoDL figure
% Figure 1

\begin{figure}[htbp]
    \centering
    \setlength{\abovecaptionskip}{0.cm}
    \includegraphics[width=8.5cm, keepaspectratio]{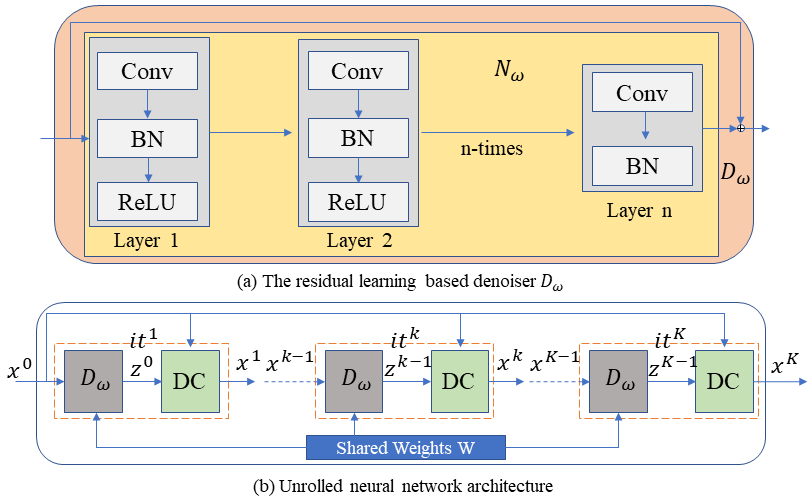}
    \caption{The originally developed fully supervised MoDL architecture \cite{aggarwal2018modl}. (a) shows the CNN based denoiser block $D_{w}\left(x\right)$. (b) is the unrolled architecture of $K$ iterations. $D_{w}\left(x\right)$ share the weights across all the $K$ iterations. 
    }
    \label{fig:1}
\end{figure}

They used fully sampled datasets to train the regularization parameters and the unrolled neural network. The final image is generated as the outputs of the neural network.

% 3 The PROPOSED METHOD
% 3.1 The Overall Framework of PARCEL
\section{The Proposed Method}
\subsection{The Overall Framework of PARCEL}
We proposed a physics-based unsupervised contrastive representation learning model PARCEL, whose overall framework is shown in Fig. 2. The specific information of co-training loss can be found in Section 3.2. Its training phase has two branches of model-based networks unfolded with the conjugate gradient algorithm for solving the following formula:
% equa 5
\begin{equation}
    x^{\ast}=\mathrm{arg}\min_{x_{j}}\left\{\frac{1}{2}\left \| A_{j}x_{j}-y \right \| _{2}^{2}+\lambda_{j} \left \| N_{w_{j}}\left ( x_{j} \right )  \right \|^{2}\right\}, j=1,2
\end{equation}
where $A_{j}=\mathbf{\Omega}_{j}FS$. Specifically, the re-undersampled mask $\mathbf{\Omega}_{j}$($j=1,2$ is the contrastive representation learning branch number), which needs to meet the following conditions: 1) two parallel networks use different selection masks, 2) the input of the network contains mostly low-frequency signals. During the training phase, two re-undersampled mask $\mathbf{\Omega}_{1}$ and $\mathbf{\Omega}_{2}$ were used to perform secondary undersampling on the original undersampled data $y$ to obtain $y_{1}$ and $y_{2}$. Thus, the input data for the contrastive learning framework is obtained, where the two parallel network has similar architecture to MoDL. Among them, the $D_{w}\left(x\right)$ module in our model is shown in Fig. 3. The parallel network is used to train the data, and the co-training loss function was specially designed to constrain the learning process of the model. During the testing phase, undersampled data are fed into the trained model to obtain representations $x_{1}$ and $x_{2}$. Then the average of $x_{1}$ and $x_{2}$ is used as the final reconstruction result.

% 3.2 The Proposed Co-training Loss
\subsection{The Proposed Co-training Loss}
We have designed a sophisticated co-training loss with three essential components to guide the two networks in capturing the inherent features and representations for MR images. Specifically, it has the undersampled calibration loss, reconstructed calibration loss and contrastive representation loss. Compared with self-supervised learning of a single network, contrastive learning of parallel network allows for more rigorous and inherent features to be captured. In this way, the network is expected to avoid learning erroneous information \cite{bardes2021vicreg}. The mathematical formula of the total co-training loss function is as follows:
% equa 6
\begin{equation}
\begin{split}
    \xi_{ct} =\frac{1}{L} ( \sum_{i=1}^{L}\ell_{uc}( Ax_{1}^{i},y^{i})+ \sum_{i=1}^{L}\ell_{uc}( Ax_{2}^{i},y^{i})) +  \frac{1}{L} ( \sum_{i=1}^{L}\ell_{rc} ( x_{1}^{i} )+ \sum_{i=1}^{L}\ell_{rc} ( x_{2}^{i})) 
\\ 
+ \frac{1}{L}\sum_{i=1}^{L}\ell_{cl} ( z_{1}^{i},z_{2}^{i} )
\end{split}
\end{equation}
where $L$ is the total number of training samples, $i$ is the $i$-th training sample, $\xi_{ct}$ represent co-training loss; $\ell_{uc}$ represent undersampled calibration loss; $\ell_{rc}$ represent reconstructed calibration loss; $\ell_{cl}$ represent contrastive representation loss; $x_{1}^{i}$ and $x_{2}^{i}$ represent the output representation of the two networks; $z_{1}^{i}$ and $z_{2}^{i}$ represent the embedding features with the respect of $x_{1}^{i}$ and $x_{2}^{i}$, respectively.

1) Undersampled Calibration Loss: The undersampled calibration loss is mainly concerned with the k-space points that have been sampled, which ensures that the reconstruction results of the sampling elements are consistent with the zero-filled k-space data from the measurement. Specifically, the output representation is undersampled by the encoding matrix $A$ and then compared with the original undersampled data. It was used to calculate the difference between the undersampled version of the network prediction and the directly zero-filled one. The specific calculation formula is as follows:
%equa 7
\begin{equation}
    \ell_{uc}\left ( Ax_{1}^{i},y^{i} \right )=\ell_{mse}\left ( F^{-1}Ax_{1}^{i},F^{-1}y^{i} \right ) 
\end{equation}

% equa 8
\begin{equation}
    \ell_{mse}\left ( a,b \right )=\frac{1}{M}\sum_{l=1}^{M}\left ( a_{l}-b_{l} \right ) ^{2} 
\end{equation}
where $F^{-1}$ represents the two-dimensional inverse Fourier transform; $M$ represents size of the input data. This operation is trying to calibrate the consistency between the undersampled portion of the output representation and the undersampled k-space data of the original input.

% network architecture
% Figure 2
\begin{figure*}[htbp]
    \centering
    \setlength{\abovecaptionskip}{0.cm}
    \includegraphics[width=11cm, keepaspectratio]{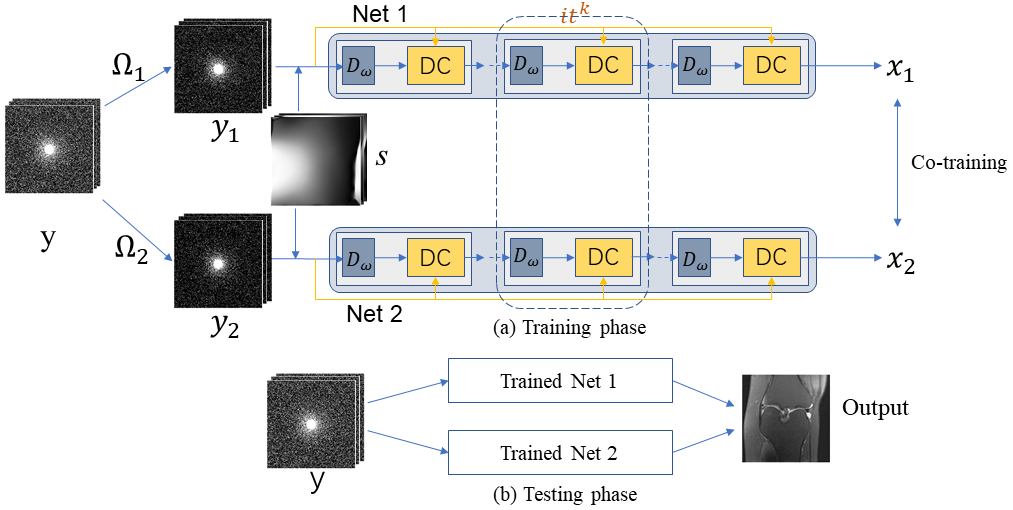}
    \caption{The overall framework of PARCEL. (a) Training phase. Reundersample  $y$ to obtain $y_{1}$ and $y_{2}$ for parallel training. The representations $x_{1}$ and $x_{2}$ are then obtained through the trained parallel network architecture. The co-training loss is specially designed to constrain the contrastive learning process. $S$ denotes sensitivity map. (b) Testing phase. Undersampled k-space data are fed into the trained model to obtain the output representation $x_{1}$ and $x_{2}$. The average of $x_{1}$ and $x_{2}$ is used as the final reconstruction result.
    }
    \label{fig:2}
\end{figure*}

2) Reconstructed Calibration Loss: The reconstructed calibration loss (obtained by applying an affine projection based on the undersampling mask) is constructed and applied to the output representation $x_{1}$ and $x_{2}$ of the parallel networks, which not only ensures that the reconstruction results do not deviate from the measurement results but also improves the signal-to-noise ratio of the reconstructed image \cite{mardani2018deep}. Specifically, the output representation is first transformed into k-space by the encoding matrix $E$ and multiplied with $\mathbf{1}-\mathbf{\Omega} $. Then the undersampled k-space data of the original input is added. Then it is transformed to the image domain using the inverse encoding matrix $E^{H}$. Finally, the mean squared error is calculated with the output representation $x$. The specific calculation formula is as follows:
% euqa 9
\begin{equation}
    \ell_{rc}\left( x\right)=\ell_{mse}\left(x,E^{H}\left (Ex\left (\mathbf{1}-\mathbf{\Omega }  \right )+y\right )\right)
\end{equation}
where $x$ represents the output representation of Net1 or Net2, $E=FS$ denotes the encoding matrix, $E^{H}=SF^{-1}$ denotes inverse encoding matrix, and  $\mathbf{1}$ denotes a matrix with the same size as $\mathbf{\Omega}$ and all elements are $1$.

3) Contrastive Representation Loss: We construct a contrastive representation loss based on contrastive representation learning, as shown in Fig. 2(a), data are expanded by generating different inputs through two transformations, and the inputs are encoded into representations $x_{1}$ and $x_{2}$. Finally, the similarity of the two output representations is maximized to ensure that the outputs of the upper and lower network are close enough. It is expected to more effectively recover high frequency information. Specifically, it's to calculate the loss between the embedding features $z_{1}$ and $z_{2}$ with the respect to the output representations $x_{1}$ and $x_{2}$. The specific calculation formula is as follows:
% equa 10
\begin{equation}
    \ell_{cl}\left ( z_{1}, z_{2} \right )=-log\frac{exp\left ( sim\left  ( z_{1}, z_{2}\right  )\right )}{exp\left ( sim\left  ( z_{1}, z_{2}\right  )\right )+\gamma  }
\end{equation}
where $z_{1}=h_{1}\left(x_{1}\right)$ and $z_{2}=h_{2}\left(x_{2}\right)$, $h_{1}$ and $h_{2}$ are the expander, which consists of a fully connected layer of size 1024 and a ReLU activation function. $sim(z_{1}, z_{2})=\frac{z_{1}^{T}z_{2}}{\left \| z_{1} \right \| \left \| z_{2} \right \| }$, $T$ represent the transpose of the matrix, $\gamma$ is a regulating parameter used to prevent the $\ell_{cl}$ from falling to 0 and $sim(z_{1}, z_{2})$ is maximized by minimizing $\ell_{cl}(z_{1}, z_{2})$ to make the two output representations $x_{1}$ and $x_{2}$ are as close as possible. The expander eliminates the information by which the two representations differ.

% Figure 3

\begin{figure}[htbp]
    \centering
    \setlength{\abovecaptionskip}{0.cm}
    \includegraphics[width=8.5cm, keepaspectratio]{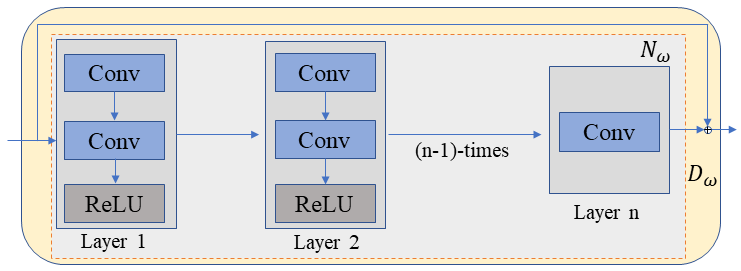}
    \caption{The denoiser $D_{w}(x)$ is used as the regularizer in this study. The batch normalization operation is replaced by the convolutional operation, which is different from its original version as shown in Fig. 1(a).
    }
    \label{fig:3}
\end{figure}

% 4 EXPERIMENTS AND RESULTS
% 4.1 Experiment Setup
% 4.1.1 Datasets
\section{Experiments and Results}
\subsection{Experiment Setup}
\subsubsection{Datasets.}

We use two datasets for evaluating the method with a public knee dataset and an in-house brain dataset. The datasets were selected based on two considerations: 1) The knee and brain MR datasets are two of the most widely used datasets in existing MR reconstruction studies. It can be regarded as a routine selection for most papers \cite{wang2021deep}, \cite{wang2020deepcomplexmri}; 2) Knee imaging and brain imaging are very important for the diagnosis of relevant diseases, including meniscus tear, bone lesion and dural thickening, etc. Therefore, reconstructing high-quality knee/brain MR images is worth of investigation. The public knee data were obtained from the NYU fastMRI database \cite{zbontar2018fastmri} and approved by the NYU School of Medicine Institutional Review Board. Fully sampled MRI data were acquired on one of three clinical 3T systems (Siemens Magneton Skyra, Prisma and Biograph mMR) or one clinical 1.5T system (Siemens Magneton Aera). Data acquisition was achieved with a 15-channel knee coil array and a conventional Cartesian 2D TSE protocol. The dataset includes data from two pulse sequences, yielding coronal proton-density weighting with (PD-FS) and without (PD) fat suppression. As per standard clinical protocol, the sequence parameters were matched as closely as possible between the two systems. The specific sequence parameters used were: echo train length 4, matrix size 320 $\times$ 320, in-plane resolution 0.5 mm $\times$ 0.5 mm, slice thickness 3 mm, and no gap between slices. The timing varied from system to system, with a repetition time (TR) of between 2200 and 3000 ms, and an echo time (TE) between 27 and 34 mm. The shape of the k-space tensor is slices $\times$ coils $\times$ height $\times$ weight. The in-house brain dataset was obtained with 3D TSE protocol by the United Imaigng system, uMR 790. The dataset contains different contrasts such as T1, T2 and PD. All data are cropped to 256 $\times$ 256. For T1-weighted images, TR = 928 ms, TE = 11 ms, voxel resolution = 0.9 $\times$ 0.9 $\times$ 0.9 mm. For T2-weighted images, TR = 2500 ms, TE = 149 ms, voxel resolution = 0.9 $\times$ 0.9 $\times$ 0.9. For PD-weighted images, TR = 2000 ms, TE = 13 ms, resolution = 1.0 $\times$ 1.1 $\times$ 1.1 mm. Informed consents were obtained from the imaging subject in compliance with the Institutional Review Board policy \cite{wang2020deepcomplexmri}. The knee dataset contains 245 volumes, and the brain dataset contains 22 volumes. The training, validation, and testing sets are divided randomly, with 6: 2: 2. In the experiment, both one and two-dimensional random undersampled masks were tested; the corresponding masks are shown in Fig. 4.

% Figure 4

\begin{figure}[htbp]
    \centering
    \setlength{\abovecaptionskip}{0.cm}
    \includegraphics[width=8.5cm, keepaspectratio]{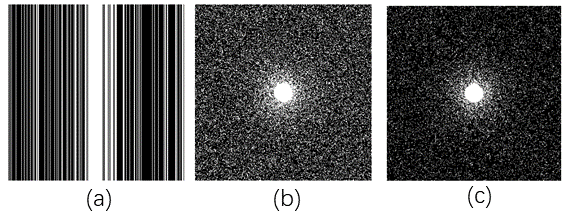}
    \caption{Two types of sampling. (a) One-dimemsional random undersampled mask with R=3. Two-dimensional random undersampled masks with (b) R=4 and (c) R=8. R represents the acceleration rate. The autocalibration signal (ACS) lines of the undersampled mask is 24.
    }
    \label{fig:4}
\end{figure}

% 4.1.2 Implementation Details
\subsubsection{Implementation Details.}
We use a deep neural network with 5 layers, each with 64 convolution kernels to achieve $D_{w}(x)$, and the size of the convolution kernel is $3\times3$, except for the last layer. Each layer contains two continuous convolution operations and a linear activation function, ReLU (rectified linear unit, $f(x)=max(0,x)$). The last layer has only one convolution operation. Among them, we have replaced BN \cite{ioffe2015batch} with the convolution operation, as shown in Fig. 3. This is similar to the practice in literature \cite{yaman2020self}. We extract the output of block $D_{w}(x)$ to the data consistency layer. In the experiment, the number of alternate iterations of network unfolding, $K$, is 5 \cite{aggarwal2018modl}. The input and output of the data consistency layer are complex values. Module $D_{w}(x)$ provides input by superimposing the real and imaginary parts in a channel. Among these, coil sensitivity is estimated from the central k-space region of each slice using the ESPIRiT \cite{uecker2014espirit} algorithm, with all assumptions known in the experiment. During training, we used ADAM \cite{kingma2014adam} optimization, and the momentum was (0.9, 0.999). The initial rate was $10^{-4}$, and the learning rate attenuation strategy was adopted \cite{li2022artificial}. The SSIM metric value of the validation set was taken as the monitoring indicator, and the learning rate was multiplied by 0.3 when the SSIM metric no longer decreased within 10 epoch periods. If the SSIM metric did not change within 50 epochs, the training ends. The total epoch was 200, and the batch size was 4. We train the network under two random masks and different acceleration factors to explore the reconstruction effect of the model under different sampling methods. The model is implemented in Pytorch and the code can be downloaded from this link: https://github.com/ternencewu123/PARCEL. 

% 4.1.3 Evaluation Metrics
\subsubsection{Evaluation Metrics.}
In the experiment, peak signal-to-noise ratio (PSNR) and structural similarity (SSIM) \cite{wang2004image} were used to quantitatively evaluate the experimental results. The SSIM index is the product of the luminance, contrast and structure measure functions. The corresponding calculation formula is:
\begin{equation}
    SSIM(\mathbf{x},\mathbf{y})=\frac{(2\mu_{x}\mu_{y}+C_{1})(2\delta_{xy}+C_{2})}
{(\mu_{x}^{2}+\mu_{y}^{2}+C_{1})(\delta_{x}^{2}+\delta_{y}^{2}+C_{2})} 
\end{equation}
where $\mu_{x}$ is the average of $x$, $\mu_{y}$ is the average of $y$, $\delta_{x}^{2}$ is the variance of $x$, $\delta_{y}^{2}$ is the variance of $y$, $\delta_{xy}$ is the covariance of $x$ and $y$. $C_{1}=(k_{1}L)^{2}$ and $C_{2}=(k_{2}L)^{2}$ are constants used to ensure stability, $L$ is the dynamic range of pixel values, $k_{1}=0.01, k_{2}=0.03$.

% 4.1.4 Compared Methods
\subsubsection{Comparision Methods.}
We compared PARCEL with five methods, SENSE \cite{pruessmann1999sense}, Variational-Net \cite{hammernik2018learning}, U-Net-256, SSDU \cite{yaman2020self} and Supervised-MoDL \cite{aggarwal2018modl} methods under different acceleration rates and sampling patterns. SENSE is a classical parallel imaging method based on coil sensitivity encoding; Variational-Net learns a variational network to accelerate MRI reconstruction, the loss function is the mean-squared error (MSE); U-Net-256 is a classical U-Net model traind in a supervised manner, where the number of channels of the last encoder layer is 256, and the loss function is the mean-squared error; SSDU is a model-based method in which an MoDL model trained in a self-supervised manner as in \cite{yaman2020self}, and the loss function is a normalized $\ell_{1}-\ell_{2}$ loss; Supervised-MoDL are trained using pairs of measurement data and ground truth images, and the loss function is the mean-squared error. The code of all methods are implemented in Python.

% 4.2 Evaluation with Different Sampling Masks
\subsection{Evaluation with Different Sampling Masks}
% 4.2.1 1D Random Sampling experiment
\subsubsection{One-Dimensional Random Sampling Results.}

To evaluate the reconstruction performance of PARCEL for the one-dimensional random sampling, the public knee dataset is used in this section, which consists of 15 coils and is complex-valued data.

% Figure 5
\begin{figure*}[htbp]
    \centering
    \setlength{\abovecaptionskip}{0.cm}
    \includegraphics[width=11cm, keepaspectratio]{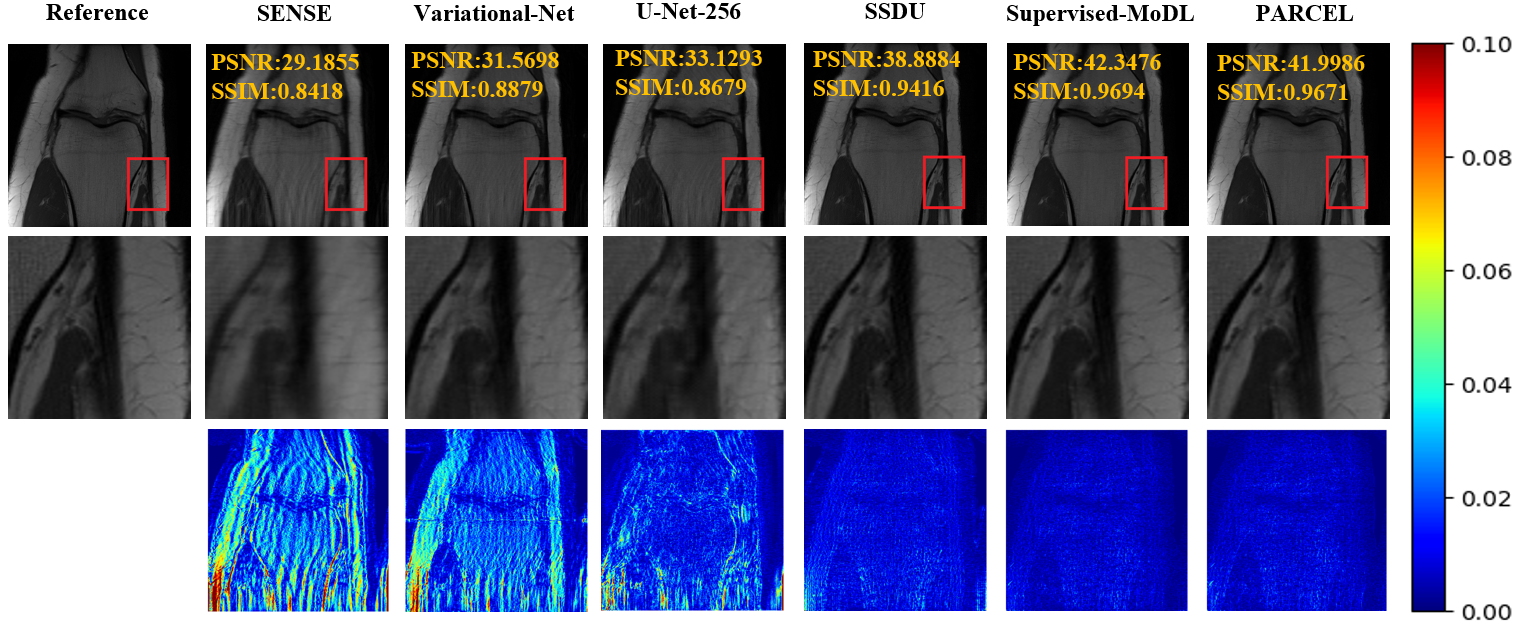}
    \caption{Reconstruction results of an example test slice from the fastMRI coronal proton density (PD) knee MRI dataset with the acceleration rate R = 3. Fully sampled images are shown in the first column for reference. The remaining columns show the reconstructed images of SENSE, Variational-Net, U-Net-256, SSDU, Supervised-MoDL and PARCEL. 
    }
    \label{fig:5}
\end{figure*}

% Figure 6
\begin{figure*}[htbp]
    \centering
    \setlength{\abovecaptionskip}{0.cm}
    \includegraphics[width=11cm, keepaspectratio]{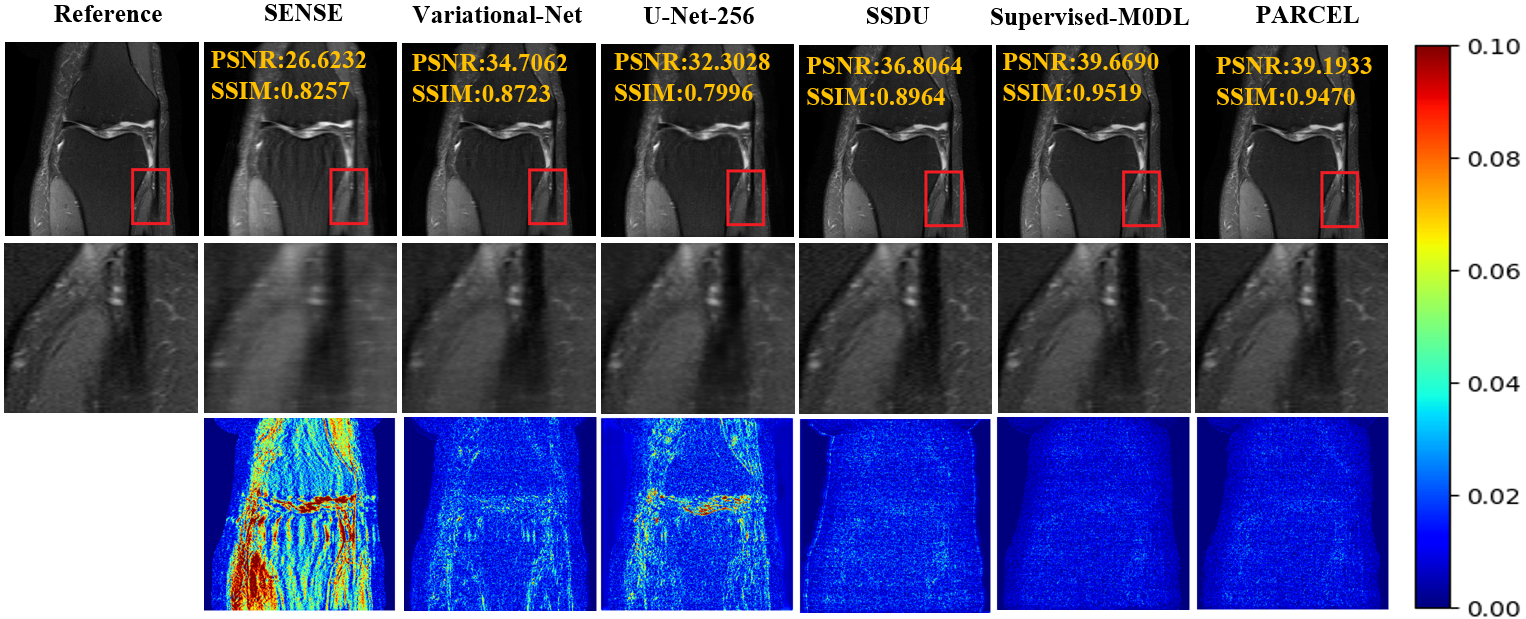}
    \caption{ Reconstruction results of an example test slice from the fastMRI coronal density weighted with fat suppression (PD-FS) dataset with the acceleration rate R=3. Fully sampled images are shown in the first column for reference. The remaining columns show the reconstructed images of SENSE, Variational-Net, U-Net-256, SSDU, Supervised-MoDL and PARCEL.
    }
    \label{fig:6}
\end{figure*}

% Figure 7
\begin{figure}[htbp]
    \centering
    \setlength{\abovecaptionskip}{0.cm}
    \includegraphics[width=8.5cm, keepaspectratio]{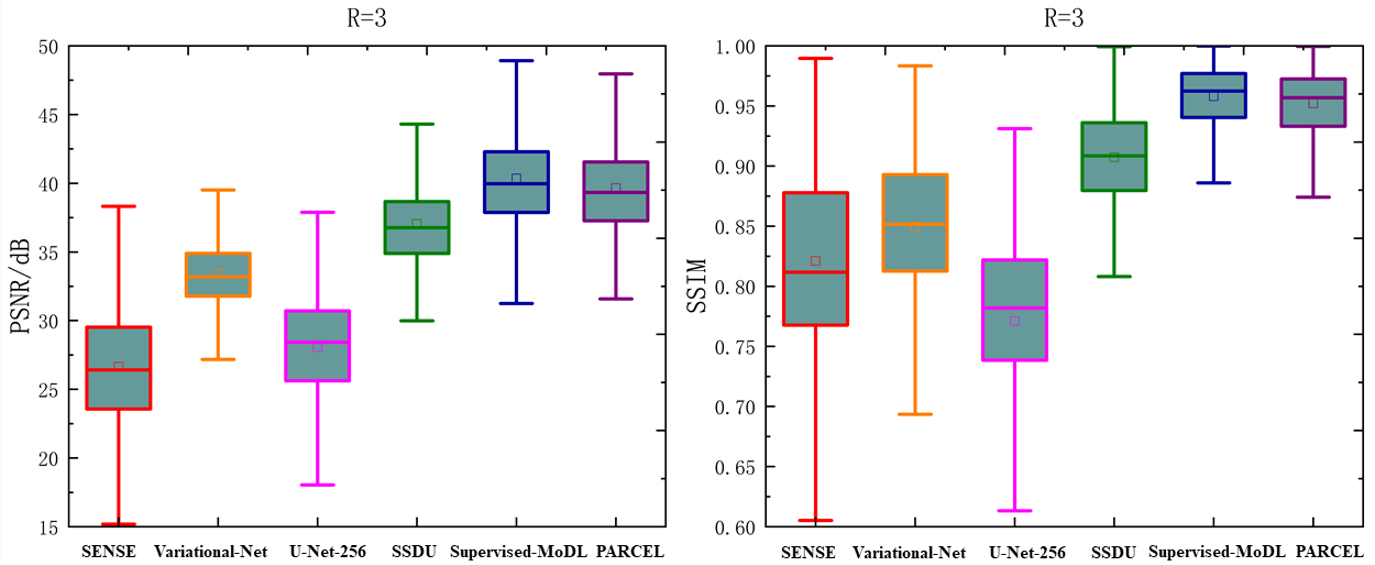}
    \caption{Box plots showing the median and interquartile range (25th-75th percentile) of PSNR and SSIM calculated using the test data with the acceleration rate R = 3. 
    }
    \label{fig:7}
\end{figure}

Fig. 5 demonstrates the reconstruction results of coronal PD images with acceleration rate of 3 using SENSE, Varia-tionalNet, U-Net-256, SSDU, Supervised-MoDL and the proposed unsupervised model PARCEL. The first row shows the reconstructed images. The second row represents local detail graphs of the reconstructed images, and the third row presents error maps using a jet color map (blue: low, red: high error) between the reconstructed and the reference images. The SENSE, Variational-Net, and U-Net-256 approaches suffer from visible residual artifacts and do not recover detailed structures due to over-smoothing problems, as shown in the error maps. SSDU, Supervised-MoDL, and our approach PARCEL remove residual artifacts and recover detailed structure. Compared with SSDU, our method achieves better reconstruction results. This is mainly due to contrastive representation learning’s ability to dig deep information along with the co-training loss constraint on the network. Furthermore, the performance of our method is close to the Supervised-MoDL and may have benefited from the constraint of reconstructed calibration loss on the reconstruction result to avoid deviation from the measured data. The quantitative metrics and error maps shown in Fig. 5 are consistent with these observations. 
 
Similar results can be observed for coronal PD-FS images with acceleration rate of 3, as depicted in Fig. 6. PARCEL and Supervised-MoDL methods achieve similar performance while improving the suppression of the visible residual artifacts in SENSE, Variational-Net and U-Net-256. The quantitative evaluation results as well as the residual artifacts in the error maps also highlight these observations. In addition, compared with coronal PD-FS images, the reconstruction quality of coronal PD images improves overall, except for the Variational-Net method.
 
 In addition to qualitative comparisons of visual quality, quantitative comparisons are also highlighted. Fig. 7 show box plots displaying the median and interquartile range (25th-75th percentile) of the quantitative metrics for PSNR and SSIM with acceleration rate 3, across all test datasets for each knee sequence. For all sequences, PARCEL and Supervised-MoDL achieve similar quantitative performance for both PSNR and SSIM, significantly outperforming SENSE, Variational-Net and U-Net-256. In particularly, for the SSIM metric, the results of the values taken are relatively concentrated, which indicates the better performance of our model. Furthermore, compared to pure neural network methods, model-driven based methods have better reconstruction performances.

% 4.2.2 2D Random Sampling Experiment
\subsubsection{Two-Dimensional Random Sampling Results.}
To further evaluate model reconstruction quality, we continued our experiments using a two-dimensional random mask for the in-house brain dataset. Fig.8 and Fig. 9 show the quantitative results about PSNR and SSIM at different acceleration rates. Compared with the knee reconstruction results with one-dimensional random sampling masks, similar trends can be obtained for the brain reconstruction with two-dimensional random sampling maks. For example, the reconstruction performance of SSDU, Supervised-MoDL and PARCEL proposed in this paper is better than SENSE, Variational-Net and U-Net-256. In addition, the results generated by PARCEL are close to the fully supervised method Supervised-MoDL. The details are shown in Fig. 7, Fig. 8 and Fig. 9.

In addition to the quantitative comparison with PSNR and SSIM, the visual quality is also presented. Fig. 10 shows qualitative reconstruction results of the brain testing set using various reconstruction methods at the acceleration rates of 4 and 8 along with their error maps using a jet color map (blue: low, red: high error). These experimental results show that PARCEL can reconstruct the images with improved PSNR and SSIM. The error map clearly shows that SSDU, Supervised-MoDL and PARCEL maintain detailed information better than the other methods. The pure U-Net network fails to achieve good results. However, it is effective in realizing the reconstruction ability of the MRI image domain. Similar to the experimental results with the Knee, SENSE, Variational-Net and U-Net-256 all produced smoothed results in the brain experiment, resulting in blurred local details, as shown in the detail enlargement in Fig. 10. In addition, at higher acceleration rates, our method PARCEL outperforms SSDU in terms of detail recovery and is close to the result of Supervised-MoDL.

% Figure 8
\begin{figure}[htbp]
    \centering
    \setlength{\abovecaptionskip}{0.cm}
    \includegraphics[width=8.5cm, keepaspectratio]{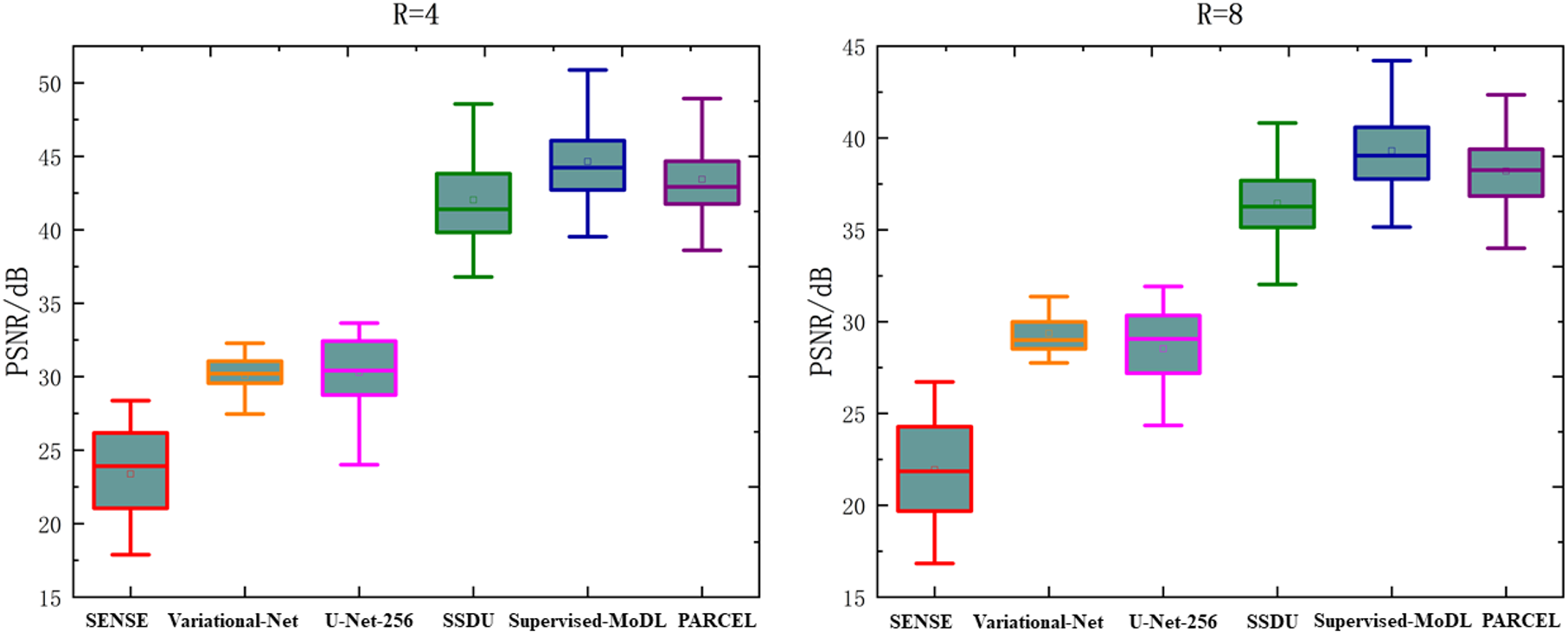}
    \caption{Box plots showing the median and interquartile range (25th-75th percentile) of PSNR calculated using the test data with the acceleration rates R = 4 and R = 8. 
    }
    \label{fig:8}
\end{figure}

% Figure 9

\begin{figure}[htbp]
    \centering
    \setlength{\abovecaptionskip}{0.cm}
    \includegraphics[width=8.5cm, keepaspectratio]{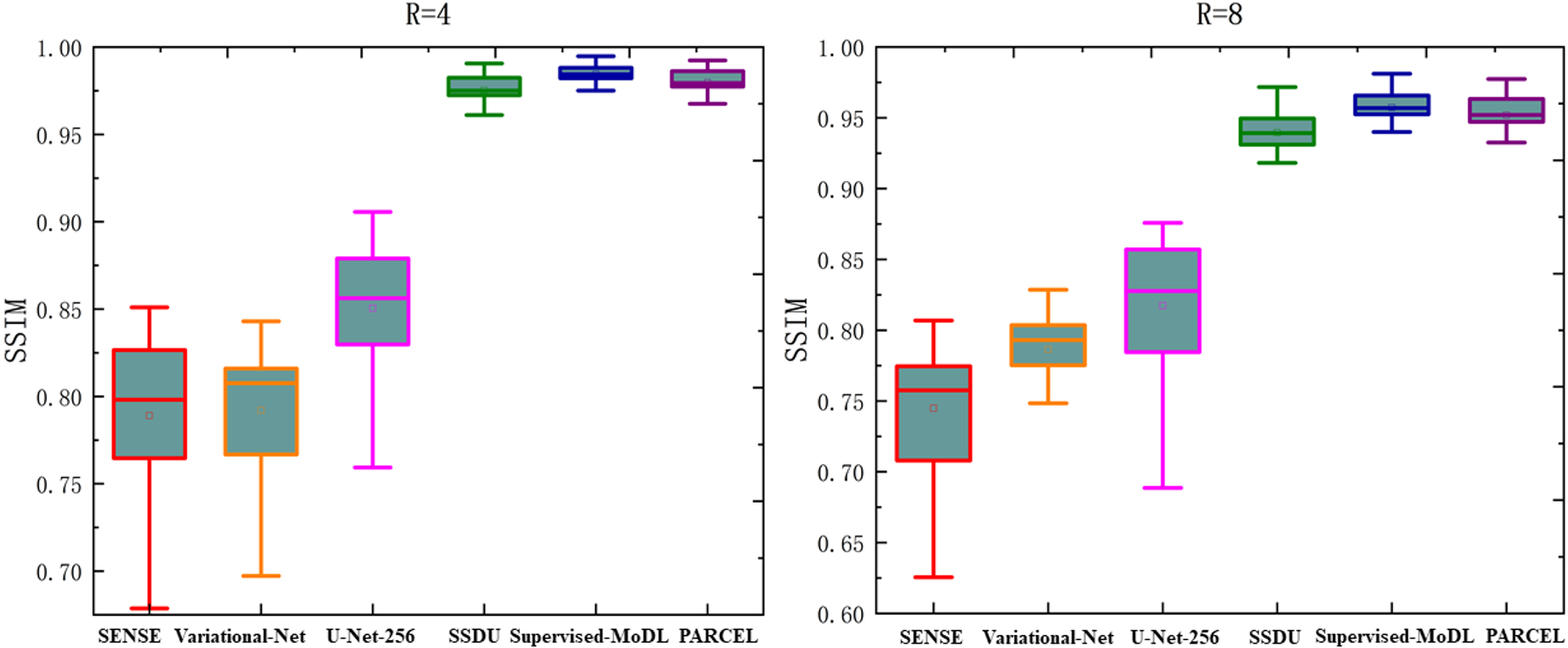}
    \caption{Box plots showing the median and interquartile range (25th-75th percentile) of SSIM calculated using the test data with the acceleration rates R = 4 and R = 8. 
    }
    \label{fig:9}
\end{figure}

% Figure 10
\begin{figure}[htbp]
    \centering
    \setlength{\abovecaptionskip}{0.cm}
    \includegraphics[width=11cm, keepaspectratio]{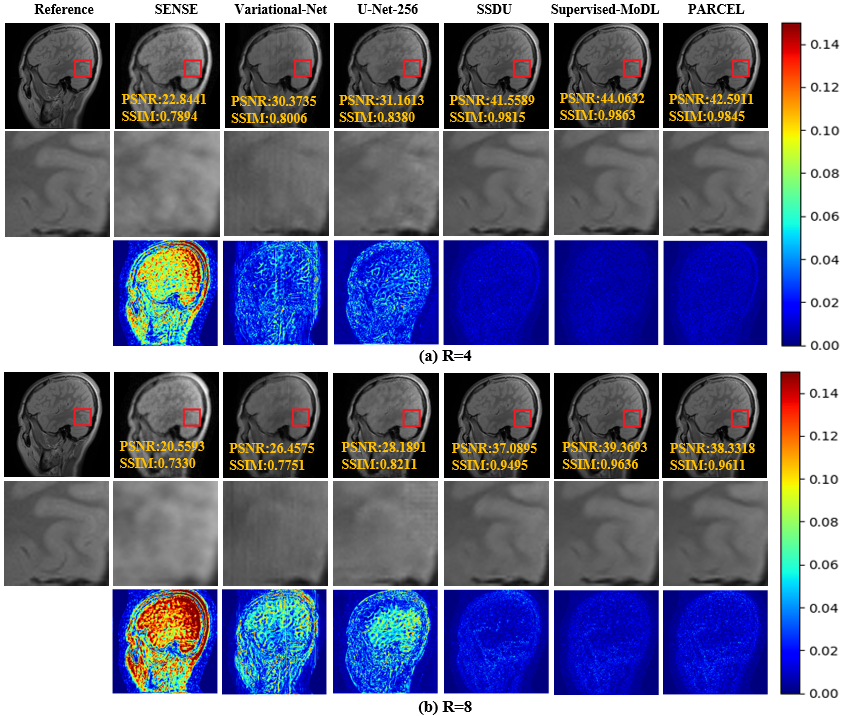}
    \caption{ Reconstruction results of an example test slice from the in-house brain dataset with acceleration rates R=4 (a) and R=8 (b). Fully sampled images are shown in the first column for reference. The remaining columns show the reconstructed images of SENSE, Variational-Net, U-Net-256, SSDU, Supervised-MoDL and PARCEL. 
    }
    \label{fig:10}
\end{figure}

%demo file is intended to serve as a ``starter %file''
%for IEEE Computer Society journal papers produced under \LaTeX\ using
%IEEEtran.cls version 1.8b and later.
%I wish you the best of success.

 

% 4.3 Validation of Contrastive Representation Learning
\subsection{Validation of Contrastive Representation Learning}
To evaluate the effects of contrastive representation learning, we conducted experiments about contrastive representation loss with the different random masks. First, experiments are performed on the one-dimensional random sampling pattern, and Fig. 11 show the quantitative comparison results of the contrastive representation loss. Among them, Single-Net is an end-to-end self-supervised MoDL model, and the loss function is the mean-squared error loss. Parallel-Net is a PARCEL model with only the undersampled calibration loss. CL is a PARCEL model with the undersampled calibration and contrastive representation loss.

From the quantitative results, it shows that the PSNR and SSIM metrics are improved after adding the contrastive representative loss. For example, the PSNR metric improved from 37.3140 dB to 39.3004 dB and the SSIM metric improved from 0.9302 to 0.9401, as shown in Fig. 11. In addition, Fig. 12 show the qualitative results. Compared with traditional self-supervised learning methods, such as Single-Net, the qualitative result of the reconstructed images has been improved by using a contrastive representative learning approach with parallel networks. Also, this phenomenon can be observed in the error map in Fig. 12.

% Figure 11 
\begin{figure}[htbp]
    \centering
    \setlength{\abovecaptionskip}{0.cm}
    \includegraphics[width=8.5cm, keepaspectratio]{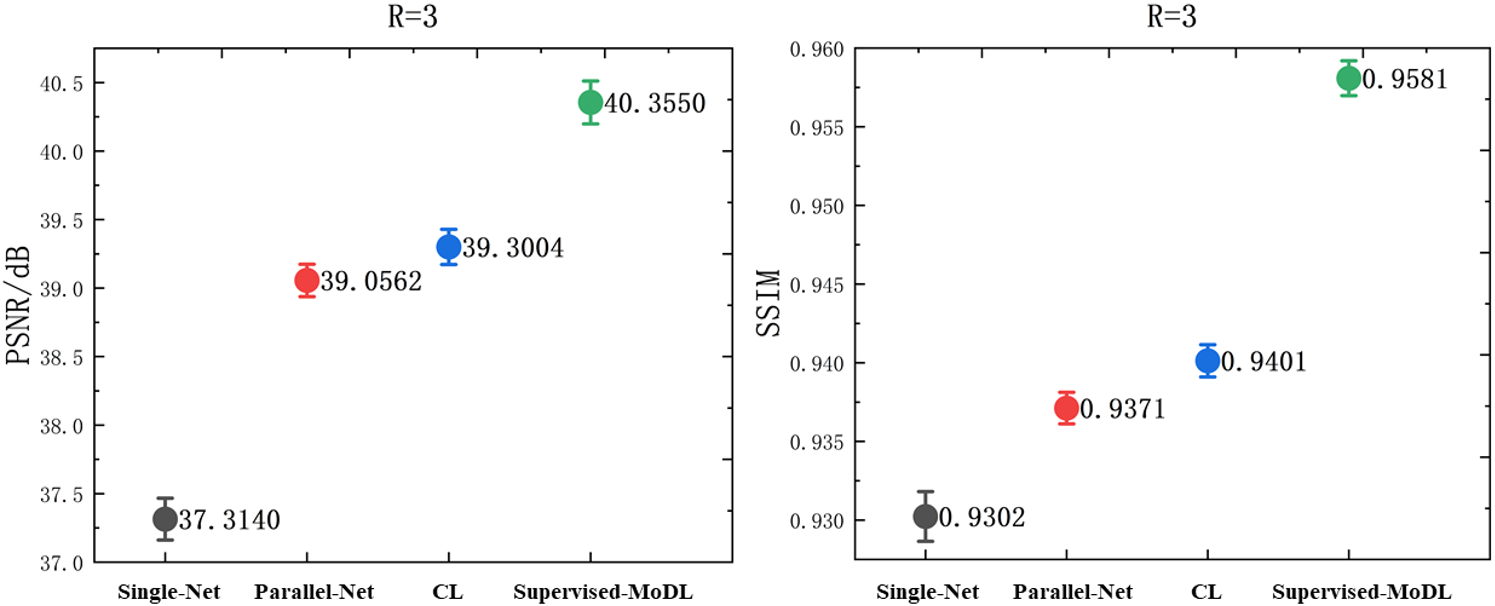}
    \caption{Quantitative comparison of contrastive representation loss about PSNR and SSIM calculated using the knee datasets with the acceleration rate of 3. The bullseye shows the means and error bars shows the standard errors on the means.
    }
    \label{fig:11}
\end{figure}

% Figure 12
\begin{figure}[htbp]
    \centering
    \setlength{\abovecaptionskip}{0.cm}
    \includegraphics[width=8.5cm, keepaspectratio]{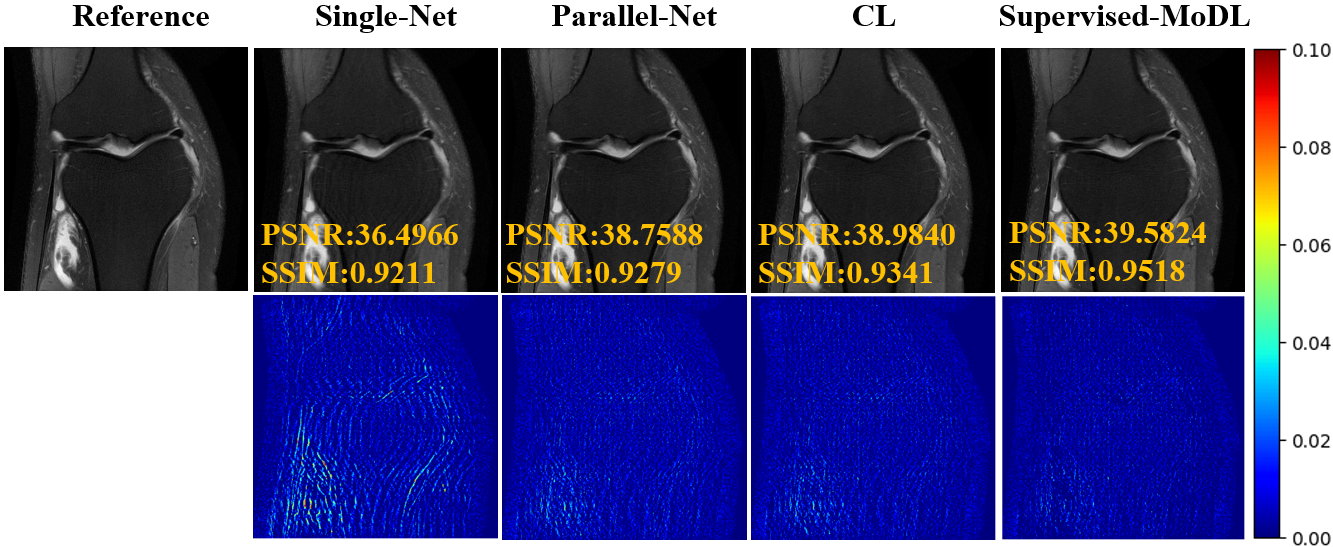}
    \caption{Reconstruction results of an example test slice from the fastMRI coronal proton density weighted with fat suppression (PD-FS) dataset with the acceleration rate 3. Fully sampled images are shown in the first column for reference. The remaining columns show the reconstructed images of Single-Net, Parallel-Net, CL and supervised-MoDL. 
    }
    \label{fig:12}
\end{figure}

In addition, we have done corresponding experiments on two-dimensional random sampling pattern. Fig. 13 and Fig. 14 show a quantitative comparison of the brain dataset with the acceleration rates of 4 and 8. From the experimental results, the quantitative results were greatly improved by adding the contrastive representation loss. For example, the PSNR metric improves from 38.5802 dB to 43.3204 dB at the acceleration rate of 4 in Fig. 13. In addition to the quantitative results, Fig. 15 show the qualitative results of the contrastive representation loss on the brain dataset. The first column indicates the reference image, and the other columns represent the reconstruction results and the corresponding error maps. Similar conclusions to the previous ones can be drawn from the results of the error maps. It is evident that contrastive representation learning improves the quality of the reconstructed image. Moreover, the detailed information of local regions is improved. Overall, the parallel network Parallel-Net achieves more information compensation than traditional self-supervised learning methods such as Single-Net, while the parallel network CL using contrastive representation learning achieves better reconstruction by learning the similarity between the high-dimensional embeddings of the upper and lower encoded representations.

% Figure 13
\begin{figure}[htbp]
    \centering
    \setlength{\abovecaptionskip}{0.cm}
    \includegraphics[width=8.5cm, keepaspectratio]{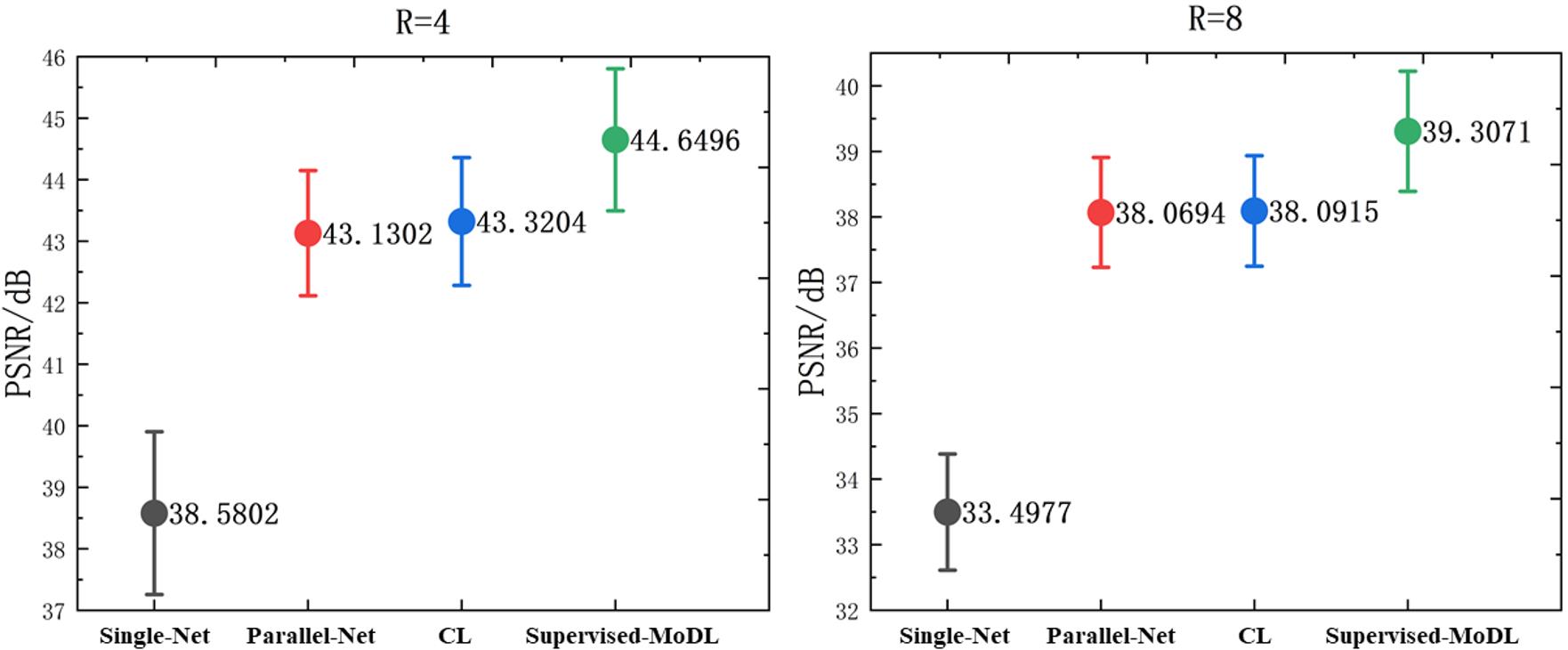}
    \caption{Quantitative comparison of contrastive representation loss about PSNR calculated using the brain dataset with the acceleration rates 4 and 8. The bullseyes represent the means and error bars show the standard errors on the means.
    }
    \label{fig:13}
\end{figure}

% Figure 14
\begin{figure}[htbp]
    \centering
    \setlength{\abovecaptionskip}{0.cm}
    \includegraphics[width=8.5cm, keepaspectratio]{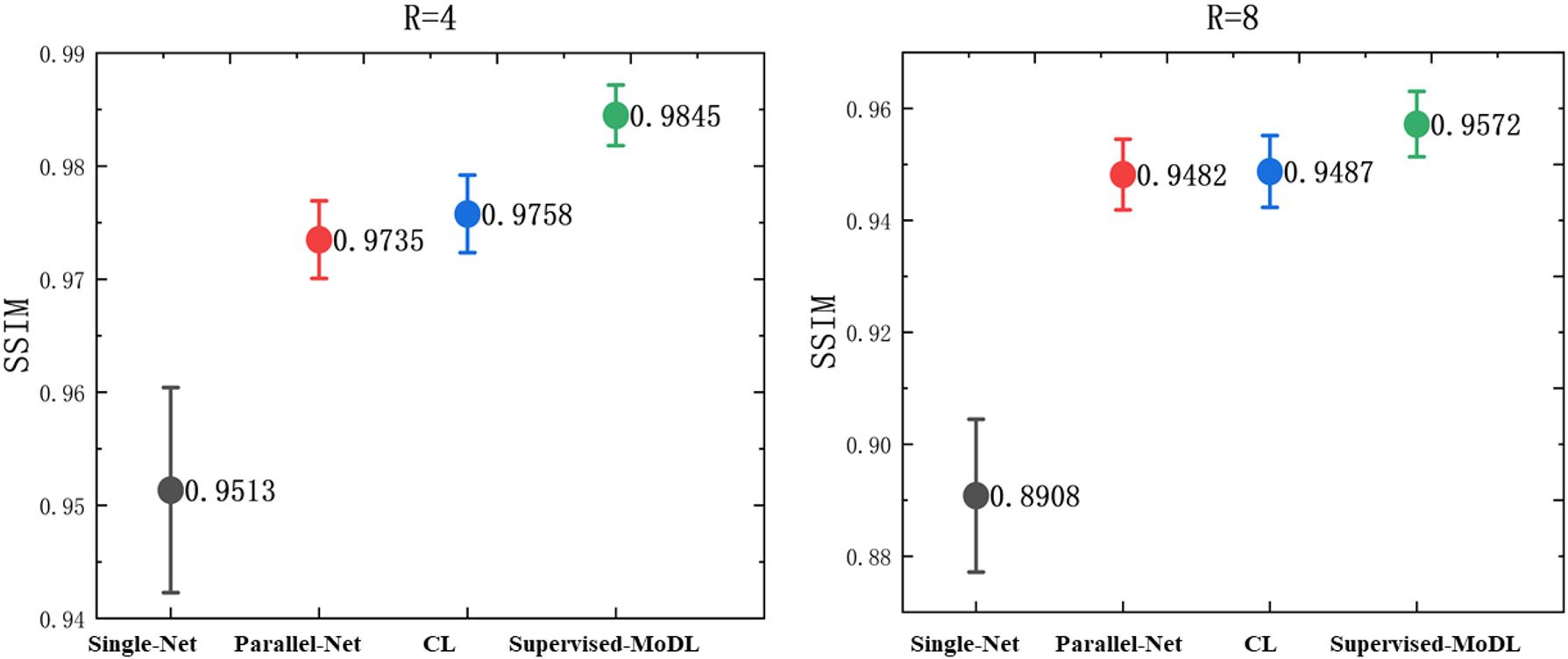}
    \caption{Quantitative comparison of contrastive representation loss about SSIM calculated using the brain dataset with the acceleration rates 4 and 8. The bullseyes represent the means and error bars show the standard errors on the means.
    }
    \label{fig:14}
\end{figure}

% Figure 15
\begin{figure}[htbp]
    \centering
    \setlength{\abovecaptionskip}{0.cm}
    \includegraphics[width=8.5cm, keepaspectratio]{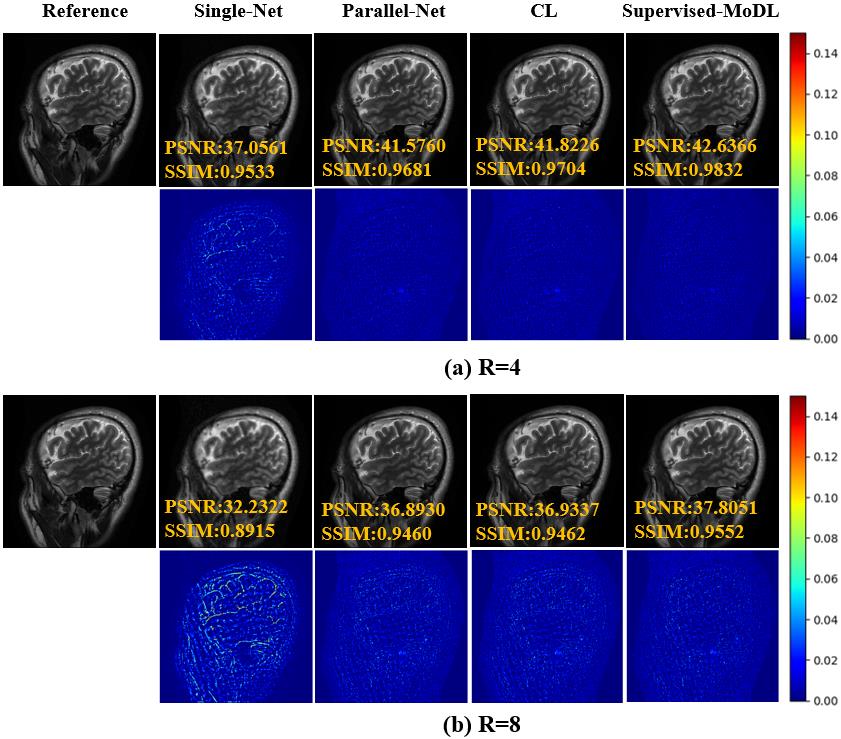}
    \caption{Reconstruction results of an example test slice from the brain dataset with the acceleration rates R=4 (a) ad R=8 (b). Fully sampled images are shown in the first column for reference. The remaining columns show the reconstructed images of Single-Net, Parallel-Net, CL and supervised-MoDL.
    }
    \label{fig:15}
\end{figure}

Meanwhile, we also compare the difference between the two outputs of the parallel network before and after the inclusion of the contrastive representation loss, and the results are shown in Table 1. According to the experimental results, after the inclusion of the contrastive representation loss, the difference between the two outputs representation of the parallel network is smaller, indicating that two output representations are closer. The reason is that contrastive representation learning mines the similar internal information of the inputs ($y_{1}$ and $y_{2}$ in Fig. 2) by learning the similarity between the high-dimensional embedding features ($z_{1}$ and $z_{2}$ in Eq. 10), thus ensuring that the output representation of the two networks is as similar as possible.

% Table 1
\begin{table}[htbp]
\caption{Quantitative Comparison Results of Parallel Network Output without $\ell_{cl}(o)$ and with $\ell_{cl}(w)$ about Contrastive Representation Loss}
\centering
\resizebox{\columnwidth}{!}{
\begin{tabular}{ccccc}
\toprule [1pt]
     Dataset & Acceleration rate & Method & PSNR/dB & SSIM  
     \\
\midrule [1pt]
    \multirow{2}{*}{Knee} &\multirow{2}{*}{R=3}& $\ell_{cl}(o)$ & 
    44.2330$\pm$2.7818 & 0.9764$\pm$0.0155 \\ 
    \multirow{2}{*}{} &  \multirow{2}{*}{} &
    $\ell_{cl}(w)$ & 45.2889$\pm$2.2453 & 0.9791$\pm$0.0135 \\  
\midrule
    \multirow{4}{*}{Brain}&
    \multirow{2}{*}{R=4} &
    $\ell_{cl}(o)$ & 50.4809$\pm$2.5057 & 0.9941$\pm$0.0015 \\  
    \multirow{4}{*}{}&
    \multirow{2}{*}{} &
    $\ell_{cl}(w)$ & 51.4517$\pm$2.8781 & 0.9962$\pm$0.0012 \\ 
    
    \multirow{4}{*}{}&
    \multirow{2}{*}{R=8} &
    $\ell_{cl}(o)$ & 45.6516$\pm$2.2844 & 0.9919$\pm$0.0024 \\  
    \multirow{4}{*}{}&
    \multirow{2}{*}{} &
    $\ell_{cl}(w)$ & 45.8389$\pm$2.3688 & 0.9938$\pm$0.0024 \\  
\bottomrule  [1pt]
\end{tabular}
}
\end{table}

%\IEEEpubidadjcol

%4.4 Ablation Study
\subsection{Ablation Study}
To evaluate the different loss function proposed, relevant experiments were carried out. Table 2 show the results of ablation study conducted with different loss functions. Among these, the calculation formula of contrastive representation loss is shown in (10), and the calculation formula of reconstructed calibration loss is shown in (9). The contrastive representation loss was previously discussed. Here, we mainly discuss the reconstructed calibration loss and the combination between reconstructed calibration loss and contrastive representation loss. The ablation experimental results are based on knee and brain datasets. For the reconstructed calibration loss, on the knee dataset, the PSNR value is 39.6380 $\pm$ 3.3139, the SSIM value is 0.9520 $\pm$ 0.0268. On the brain dataset, the PSNR value is 43.3696 $\pm$ 2.6607 and 38.1895 $\pm$ 2.1416 , the SSIM value is 0.9795 $\pm$ 0.0083 and 0.9514 $\pm$ 0.0161, when the acceleration rates are 4 and 8, respectively. Similarly, for the combination between reconstructed calibration loss and contrastive representation loss, on the knee dataset, the PSNR value is 39.6476 $\pm$ 3.3136, the SSIM value is 0.9521 $\pm$ 0.0266. On the brain dataset, the PSNR value is 43.4399 $\pm$ 2.6598 and 38.2024 $\pm$ 2.1305 , the SSIM value is 0.9797 $\pm$ 0.0083 and 0.9519 $\pm$ 0.0161, respectively. 

From the PSNR and SSIM metrics, the reconstruction quality has been improved to a certain extent. The results on the knee and brain datasets show that the combination between reconstructed calibration loss and contrastive representation loss achieve the best result. In general, the reconstruction calibration loss between parallel network can further improve the reconstruction performance of the model. In addition, the ability of contrastive representation learning to extract deep information improve the reconstruction effect. 

% Table 2

\begin{table*}[htbp]

\caption{Quantitative Comparison of Reconstructed Magnetic Resonance Imaging Using PARCEL with the Different Losses and Two Baseline Models with Two Acceleration Rates (MEAN $\pm$ STD)}
\centering
\resizebox{\textwidth}{!}{
\tiny % 控制字体大小
\begin{tabular}{ccccc}
\toprule [1pt]
     Dataset & Acceleration rate & Method & PSNR/dB & SSIM  
     \\
\midrule [1pt]
    % Single-Net
    \multirow{5}{*}{Knee} &\multirow{5}{*}{R=3}
    & Single-Net & 37.3140$\pm$3.3727 & 0.9302$\pm$0.0349 
    \\
    %PARCEL(luc)
    \multirow{5}{*}{} &  \multirow{5}{*}{} &
    PARCEL($\ell_{uc}$) & 39.0562$\pm$2.6119 & 0.9371$\pm$0.0221 \\  
    % PARCEL(luc+lcl)
    \multirow{5}{*}{} &  \multirow{5}{*}{} &
    PARCEL($\ell_{uc}+\ell_{cl}$) & 39.3004$\pm$2.8372 & 0.9401$\pm$0.0226 
    \\ 
    % PARCEL(luc+lrc)
    \multirow{5}{*}{} &  \multirow{5}{*}{} &
    PARCEL($\ell_{uc}+\ell_{rc}$) & 39.6380$\pm$3.3139 & 0.9520$\pm$0.0268 
    \\ 
    % PARCEL(luc+lcl+lrc)
    \multirow{5}{*}{} &  \multirow{5}{*}{} &
    PARCEL($\ell_{uc}+\ell_{cl}+\ell_{rc}$) & 39.6476$\pm$3.3136 & 0.9521$\pm$0.0266 
    \\
    
\midrule
    \multirow{10}{*}{Brain}&
    \multirow{5}{*}{R=4} &
    Single-Net & 38.5802$\pm$3.3419 & 0.9513$\pm$0.0229 
    \\  
    \multirow{10}{*}{}&
    \multirow{5}{*}{} &
    PARCEL($\ell_{uc}$) & 43.1302$\pm$2.5712 & 0.9735$\pm$0.0087 \\
    \multirow{10}{*}{} &  \multirow{5}{*}{} &
    PARCEL($\ell_{uc}+\ell_{cl}$) & 43.3204$\pm$2.6267 & 0.9758$\pm$0.0087
    \\
    %PARCEL(luc+lrc)
    \multirow{10}{*}{} &  \multirow{5}{*}{} &
    PARCEL($\ell_{uc}+\ell_{rc}$) & 43.3696$\pm$2.6607 & 0.9795$\pm$0.0083 
    \\
    \multirow{10}{*}{} &  \multirow{5}{*}{} &
    PARCEL($\ell_{uc}+\ell_{cl}+\ell_{rc}$) & 43.4399$\pm$2.6598 & 0.9797$\pm$0.0083 
    \\ 
    % R=8
    \multirow{10}{*}{}&
    \multirow{5}{*}{R=8}&
    Single-Net & 33.4977$\pm$2.2391 & 0.8908$\pm$0.0345 
    \\
    \multirow{10}{*}{}&
    \multirow{5}{*}{}&
    PARCEL($\ell_{uc}$) & 38.0694$\pm$2.1187 & 0.9482$\pm$0.0160 \\
    \multirow{10}{*}{}&
    \multirow{5}{*}{}&
    PARCEL($\ell_{uc}+\ell_{cl}$) & 38.0946$\pm$2.1319 & 0.9487$\pm$0.0162
    \\
    \multirow{10}{*}{}&
    \multirow{5}{*}{}&
    PARCEL($\ell_{uc}+\ell_{rc}$) & 38.1895$\pm$2.1416 & 0.9514$\pm$0.0161 
    \\
    \multirow{10}{*}{}&
    \multirow{5}{*}{}&
     PARCEL($\ell_{uc}+\ell_{cl}+\ell_{rc}$) & 38.2024$\pm$2.1305 & 0.9519$\pm$0.0161 
    \\
    
\bottomrule [1pt]
\end{tabular}
}
\end{table*}

% 5 DISCUSSIONS
\section{Discussions}
The experimental results show that the MRI reconstruction model constructed by unsupervised contrastive representation learning outperform the classical parallel imaging algorithm and a newly proposed self-supervised method \cite{yaman2020self} with improved model stability while gradually approaching the performance of the supervised method. Compared with the pure neural network model, the iteratively unfolded network achieves better results per the comparison between the U-Net-256 and PARCEL model shown in Fig. 5 and Fig. 6.

In this study, we used a 5-layer deep neural network to realize the denoising module, as shown in Fig. 3. This part of the neural network has no fixed requirements, but can utilize the classical U-Net. The reconstruction results might be improved by using more complex network structures. In addition, the sharing of network parameters allows the network to perform a greater number of iterations without increasing the number of parameters. Other optimization methods, such as ISTA \cite{beck2009fast}, can be employed to solve the optimization problem in (2). In the future work, we can investigate new optimization methods to solve the problems in (2). In addition, there are many effective solutions for dealing with complex-valued data \cite{feng2021dual}, \cite{feng2021donet}, \cite{wang2020deepcomplexmri}. Our framework is flexible regarding the networks to be utilized. Therefore, these solutions can be added to further enhance the performance of our proposed method. In addition, in this study, reconstructed multi-coil MR images and corresponding error maps were plotted to qualitatively evaluate the performance. Quantitative evaluations were conducted by calculating objective metrics, including PSNR and SSIM. In our following study, we may consider some subjective indicators provided by radiologists, such as SaMDs \cite{machal2022impact}, to investigate the clinical acceptance of the reconstruction images. 

Furthermore, regarding the generalization of the algorithm, the main concern is whether it works well for data that has not been seen before. Compared with existing methods that require supervised data, our unsupervised approach can be more easily adapted to new data distributions in real-word applications, and can be fine-tuned online for undersampled data acquired to ensure the performance of the model. In addition, the algorithm proposed in this paper is better in terms of explainability compared to the pure deep neural network approach. The proposed model is solved by using the conjugate gradient-based optimization algorithm and unrolled into neural networks, and thus there exists mathematical explanations. From the MR imaging point of view, the data fidelity term is a prerequisite used to reconstruct faithfully the images, which can better help the radiologists visualize the interested regions and detect possible diseases.

% 6 CLUSION
\section{Conclusion}
In this paper, we propose a physics-based unsupervised contrastive representation learning method to speed up parallel MR imaging. It has a parallel framework to contrastively learn two branches of model-based unrolling networks directly from augmented undersampled multi-coil k-space data. To guide the two networks in capturing the inherent features and representations for MR images, a sophisticated co-training loss with three essential components has been designed. Finally, the final MR image is reconstructed with the trained contrastive networks. PARCEL was evaluated on two vivo datasets and compared to five state-of-the-art methods. The results show that PARCEL can learn useful representations for more accurate MR reconstructions without relying on fully sampled datasets. In the future we will investigate contrastive representation learning for dynamic or multi-contrast MR imaging.


\section*{Acknowledgments}This research was partly supported by Scientific and Technical Innovation 2030-"New Generation Artificial Intelligence"  Project (2020AAA0104100, 2020AAA0104105), the National Natural Science Foundation of China (61871371),  Guangdong Provincial Key Laboratory of Artificial Intelligence in Medical Image Analysis and Application (Grant No. 2022B1212010011), the Basic Research Program of Shenzhen (JCYJ20180507182400762), Shenzhen Science and Technology Program (Grant No. RCYX20210706092104034), Youth Innovation Promotion Association Program of Chinese Academy of Sciences (2019351), AND the Key Technology and Equipment R$\And $D Program of Major Science and Technology Infrastructure of Shenzhen: 202100102 and 202100104.

\clearpage
\bibliographystyle{unsrt}
\bibliography{main}
\end{document}